A note on the policy implications of the fiscal multiplier

Evangelos F. Magirou
Athens University of Economics and Business
Email: efm@aueb.gr
Web page:  http://www.aueb.gr/users/magirou

1. Introduction

In the wake of the unexpectedly deep recession in Greece that followed the austerity policies agreed upon between the country, the IMF and the European Union [IMF 2012, 2013a,b], a public discussion started on whether an alternative, less restrictive economic policy would have had better results.  This discussion became even more heated following the publication of the IMF report by Blanchard and Leigh [2013], where the GDP – government surplus multiplier was re-estimated upwards.  It was argued in the press, see for instance the article by H. Schneider in the Washington Post [2013], that the drop in Greece's GDP projected in the IMF's Debt Sustainability Calculations [IMF 202, 2013a,b] was in error, since it relied on too low a value of the multiplier.

This note is an analysis of the dynamical aspects of the GDP / government surplus multiplier.  Several expressions for the multiplier appear in the literature in terms of more fundamental macroeconomic parameters - see Woodford [2010] or the Wikipedia article on the Fiscal Multiplier for a summary presentation.  We will assume that the multiplier is constant, exogenously given, and we will focus on its policy implications, namely whether the value of the multiplier has any relevance on the choice of government budget policy.  We will show the (at first) counterintuitive result that in order to reduce Debt/GDP, countries with high Debt to GDP should go into further debt, as long as the Debt to GDP ratio is roughly greater than the inverse of the  multiplier.  Thus small values of the multiplier make further debt undesirable, and vice versa.

The models used here are simple and the result is essentially an exercise in economic dynamics.  However the policy implications are of some importance as long as any type of multiplier effect is valid.

2. One period analysis

Assume that the total debt of a country at the end of year t is $D_t$, while the same year's GDP is $G_t$ (in nominal terms).  We assume that the entire debt is in the national currency.  If the interest rate during year t is $r_t$, and $X_t$ are the debt payment we have by definition

$$D_t = D_{t-1}(1+r_t) - X_t \qquad (1)$$

If the GDP deflator is $p_t$ and $g'_t$ the real GDP growth we have

$$G_t = G_{t-1}(1+g'_t)(1+p_t) = (1+g_t)G_{t-1} \qquad (2)$$

where $g_t = (1+g'_t)(1+p_t) - 1$.

Denoting by $d_t$ the ratio $D_t/G_t$ (debt to GDP ratio), dividing (1) by (2) and setting $x_t = X_t/G_t$ we have the equation

$$d_t=d_{t-1}(1+r_t)/(1+g_t)-x_t \qquad (3)$$

The interest and capital payments $X_t$ arise from government primary surpluses, income from privatizations and other sources, including autonomous debt reductions.

The concept of the multiplier between GDP and government surplus is used to capture the recessionary effect of budget surpluses. Thus if $x_t^{nom}$, $g_t^{nom}$ are the nominal scenario assumptions of some economic plan, we assume that a change $\Delta x_t$ in the nominal surplus will cause a change in growth $\Delta g_t = -\eta \Delta x_t$, $\eta$ being the fiscal multiplier (expected to be a positive number). Hence a change $\Delta x_t$ in primary surplus will cause a GDP growth of $g_t = g_t^{nom} - \eta \Delta x_t$.

To see the total effect of a change in primary surplus, look at first order changes, i.e. assume that the g's are small compared to 1 (an exact analysis would give essentially the same results). In that case the fundamental dynamics in (3) simplify to

$$d_t=d_{t-1}(1+r_t-g_t) - x_t \qquad (4)$$

Setting $x_t = x_t^{nom} + \Delta x_t$ and hence $g_t = g_t^{nom} - \eta \Delta x_t$ we have
$$d_t = d_{t-1}(1+ r_t - g_t^{nom} + \eta \Delta x_t) - x_t^{nom} - \Delta x_t \text{ or}$$

$$d_t = d_{t-1}(1+ r_t - g_t^{nom}) - x_t^{nom} + (\eta d_{t-1} - 1)\Delta x_t \qquad (5)$$

To better see the effect of changes in primary surplus consider the dynamics in (4) as a nominal debt trajectory $d_t^{nom}$ which satisfies

$$d_t^{nom} = d_{t-1}^{nom}(1+r_t - g_t^{nom}) - x_t^{nom} \qquad (4')$$

Subtracting (4') from (5) and setting $\Delta d_t = d_t - d_t^{nom}$ we have the dynamics

$$\Delta d_t = \Delta d_{t-1}(1+ r_t - g_t^{nom}) + (\eta d_{t-1} - 1)\Delta x_t \qquad (6)$$

The last equation and in particular its last term shows that in case the coefficient of the change in primary surplus (with respect to some nominal scenario) is positive, the primary surplus should decrease in order to decrease the debt to GDP ratio for the current year, and vice versa. Thus, primary surpluses should decrease in case $\eta d_{t-1} > 1$ or equivalently $d_{t-1} > 1/\eta$. For multipliers of the order of 1, and debts well over 100%, a decrease of primary surpluses seems to be in order. Conversely, small multiplier values, e.g. 80% make the debt criterion more stringent.

3. Multiperiod analysis

The impact of a change $\Delta x_t$ in the primary surplus from period t to period $t+\tau$ assuming no changes in the intervening periods is the change at t $\Delta d_t = (\eta d_{t-1}^{nom} - 1)\Delta x_t$ (assuming that $d_{t-1}^{nom} = d_{t-1}$) multiplied by the product of the factors $(1+ r_{t+j} - g_{t+j}^{nom})$, that is

$$\Delta d_{t+\tau} = (\eta d_{t-1}^{nom} - 1)\Delta x_t \prod_{j=1}^{\tau}(1 + r_{t+j} - g_{t+j}^{nom})$$

In case there are changes in the primary surpluses in future periods, to first order these effects are additive and thus

$$\Delta d_{t+\tau} = \sum_{m=1}^{t+\tau}(\eta d_{m-1}^{nom} - 1)\Delta x_m \prod_{j=m-t}^{\tau}(1 + r_{t+j} - g_{t+j}^{nom}) \qquad (7)$$

4. A numerical example

Consider the following dynamics $r_t=3\%$, $g_t=2\%$ for all $t$, $d_0=100$ and $x^{nom}_t=2$ for all t, i.e. a country with a debt of 100% its GDP, a growth rate of 2%, and debt interest rate 3%. A direct calculation using (3) gives a debt to GDP in the tenth year equal to $d_{10}= 89,34$. We take a fiscal multiplier equal to 2 and examine a change in the surplus $\Delta x_1=1$. Then a direct calculation results in $d_{10}= 90,45$ which a change $\Delta d_{10}=1,11$. The first order approximation derived above in (6) is $(2*100-100)*((1+3\%)/(1+2\%))^9 = 1,09$ which is accurate to first order.

Similarly a surplus change in the fourth period equal to $\Delta x_4 =-1$ results in $d_{10}= 88,40$ i.e. $\Delta d_{10}= -0,94$. The approximation in equation (7) gives $(2*96,9-100)*(-1)(1,03/1,02)^5=-0,98$, since $d^{nom}_4=96,9$. Finally a change of the above magnitudes at both times 1 and 4 gives after direct calculation $d_{10}=89,50$ with a change $0,16$ which is approximately the sum of the differences $1,09$ and $-0,94$, as stated in equation (7).

5. Conclusion

Although economic policy considerations, especially in crisis times, are perhaps too complicated to be amenable to quantitative analyses, the above remarks might be of interest in order to assess small changes in policies that are necessary to fine tune agreements like those of the Memorandum between the IMF, the European Union and Greece. In such a case, the strong assumptions used in the above analysis might be plausible.

6. References

*All internet references were accessed in October 2013.*